\documentclass[prb,twocolumn,showpacs]{revtex4}
\usepackage{hyperref,graphicx,dcolumn}
\usepackage{amsfonts,amsmath, amssymb,bm}

\newcommand{\epsdir}{./}

\newcommand{\myFig}[5]{ %
\begin{figure}[htb] 
\begin{center} 
\includegraphics[width=#1\columnwidth,height=#2\columnwidth,clip=true]{\epsdir/#3}
\caption{#4} \vspace{-0.5cm} \label{#5} 
\end{center} \end{figure}}

\begin{document}

\title{Current-driven magnetic rearrangements in spin-polarized point contacts}
\author{Maria Stamenova, Stefano Sanvito} \email[Contact email address: ]{sanvitos@tcd.ie}
\affiliation{Department of Physics, Trinity College, Dublin 2, Ireland} 
\author{Tchavdar N. Todorov}
\affiliation{School of Mathematics and Physics, Queen's University of Belfast, Belfast BT7 INN, UK}

\begin{abstract}
A new method for investigating the dynamics of atomic magnetic moments in current-carrying magnetic point contacts under bias is presented. This combines the non-equilibrium Green's function (NEGF) method for evaluating the current and the charge density with a description of the dynamics of the magnetization in terms of quasistatic thermally-activated transitions between stationary configurations. This method is then implemented in a tight-binding (TB) model with parameters chosen to simulate the main features of the electronic structures of magnetic transition metals. We investigate the domain wall (DW) migration in magnetic monoatomic chains sandwiched between magnetic leads, and for realistic parameters find that collinear arrangement of the magnetic moments of the chain is always favorable. Several stationary magnetic configurations are identified, corresponding to a different number of Bloch walls in the chain and to a different current. The relative stability of these configurations depends on the geometrical details of the junction and on the bias, however we predict transitions between different configurations with activation barriers of the order of a few tens of meV. Since different magnetic configurations are associated to different resistances, this suggests an intrinsic random telegraph noise at microwave frequencies in the $I$-$V$ curves of magnetic atomic point contacts at room temperature. Finally, we investigate whether or not current induced torques are conservative.
\end{abstract}


\maketitle
\section{Introduction}

Most of the intriguing properties of ferromagnetic nanoscale atomic structures arise from the close interplay between magnetic phenomena and electronic transport. As the magnetization can be controlled on a length scale smaller that the spin diffusion length of the conduction electrons \cite{bruno}, the spin scattering is affecting the overall resistance of an atomic ferromagnetic device. This is the principle behind the giant magnetoresistance effect (GMR) \cite{baibich,binasch}. Remarkably also the opposite effect is possible, i.e. the electronic current, as proposed by Slonczewski \cite{slon}, can \textit{transfer spin} and alter the magnetic configuration of the underlying ferromagnetic structure. Magnetization switching, caused by spin-polarized currents, has been observed experimentally in point contact measurements \cite{rippard,chen}, and in nanopillars \cite{urazhdin}. 

It is then clear that the modeling of these atomic-scale ferromagnetic devices requires the combined description of electronic transport and of the magnetization dynamics at the atomic level. For this purpose we have developed a general scheme for evaluating spin-currents and associated current-induced torques, which allows us to investigate the magnetization dynamics and the transport of magnetic point contacts under bias. Our problem and our method mimic closely, in philosophy, electromigration problems (thermally activated current-driven structural rearrangements), where now the direction of the local magnetic moments takes the place of the atomic positions as ``reaction coordinate''. 

Although our scheme is general and is conceptually transferable to first-principles Hamiltonians (for instance, within density functional theory), here we apply the method to a simple self-consistent tight-binding (TB) model. This has the benefit of being reasonably realistic while keeping the computational overheads to a minimum.    

The paper is organized as follows: in the next section we introduce our method, describe the model and sketch out the techniques, used for the calculations. Then, in section II.C., we discuss our approach to the interplay between spin-polarized transport and the magnetic configuration, applied to a particular atomic structure. In section III we report a set of results, which explore the stability and the activation barriers for transitions between various MM configurations under bias, as well as the effect of the model parameters on the physical properties of the system. Finally, we carry out a numerical test to see whether or not the torques in these open-boundary non-equilibrium systems are conservative.

\section{The Method}

\subsection{General Idea}

Our scheme for studying current-induced dynamical effects of the magnetization in atomic-sized nanostructures is a generalization of the combined quantum-classical dynamical methods used in electromigration problems \cite{ch_frc}. Here we treat the magnetic degrees of freedom as classical variables and the conduction electrons as a quantum system. This is appropriate when the magnetic moment (MM) arises from some deep orbital levels, such as in the case of rare earth ferromagnets, but it may appear questionable for magnetic transition metals (Fe, Co and Ni), where the $d$ electrons responsible for the moment also take part in the conduction \cite{mazin}. However, since the Coulomb energy is orders of magnitude larger than any energies connected with the electron flow, it is safe to assume that only the direction of the local atomic MM is affected by the current but not its magnitude. This effectively is an adiabatic approximation, in the spirit of the Born-Oppenheimer approximation for the nuclear dynamics, where now the orientation of the local MMs is a slow variable compared with the internal electron-electron interactions that set the magnitude of the MMs \cite{antropov1,antropov2}. The Hamiltonian for the combined conduction electron/MM system can be then written in general as
\begin{equation} \label{Ham}
H(\{\phi\})=H_\mathrm{e}+V_\mathrm{es}(\{\phi\})\;,
\end{equation}
where we have isolated the interaction of the conduction electrons with the local MM, 
$V_\mathrm{es}(\{\phi\})$, from the ``free'' electron Hamiltonian $H_\mathrm{e}$. 
In this framework the local moments are uniquely specified by a set of angles $\{\phi\}$ with 
respect to a given direction.

We may now write down the generalized forces (in this case, torques) conjugate to the classical variables $\{\phi\}$:
\begin{equation} \label{hft}
T=-\langle\Psi|\frac{\partial H(\{\phi\})}{\partial \phi}|\Psi\rangle
\end{equation}  
where $|\Psi\rangle$ is a state vector of the electronic system. Equation (\ref{hft}) has the appearance of the usual Hellmann-Feynman theorem for stationary states. However it is valid in general dynamical situations, for systems driven arbitrarily far from equilibrium \cite{ventra,ch_tdtb}. 

The set of equations (\ref{Ham}) and (\ref{hft}), combined with an appropriate method for calculating the non-equilibrium electron state vector $|\Psi\rangle$, and therefore the current, is the basis for our method for describing the interplay between transport and magnetic properties. In this work, we seek to map out the activation energy barriers for magnetic rearrangements, in order to determine the preferential magnetic configurations of the system and to study transitions between them. We achieve that as follows. First, we seek the stable configurations. We evaluate the non-equilibrium state vector $|\Psi\rangle$, in a one-electron picture,  for a given starting MM configuration $\{\phi^0\}$ by solving the scattering problem associated with the Hamiltonian $H(\{\phi^0\})$. Then, by using equation (\ref{hft}) the torques for that configuration are calculated. Then static relaxation of the torques is carried out (by steepest descent), while recalculating the self-consistent current-carrying electronic structure, and torques, at every step. The procedure continues until all torques vanish. This condition gives the stationary magnetic configurations $\{\Phi\}$.

Once the stationary magnetic configurations have been found, we can calculate the activation energy barriers for thermally activated transitions between two different configurations $\{\Phi^\mathrm{initial}\}$ and $\{\Phi^\mathrm{final}\}$. We choose a single classical dynamical variable $\phi_{j}$ as the reaction coordinate and rotate it its starting value $\phi_{j}^\mathrm{initial}$ to its final value $\phi_{j}^\mathrm{final}$. At every step on the way the torques acting on all other MMs are kept relaxed to zero. The work done by the classical degrees of freedom during this quasistatic transition is then obtained by integrating the torque on the reaction coordinate $\phi_{j}$ over the migration path. The work done over the full transition is
\begin{equation} \label{dW}
W=-\int_{\Phi^\mathrm{initial}}^{\Phi^\mathrm{final}} T_j {\rm{d}}{\phi_j}\:.
\end{equation} 
The energy barrier profile, on the other hand, is given by
\begin{equation} 
W(\Phi_j)=-\int_{\Phi^\mathrm{initial}}^{\Phi_j} T_j {\rm{d}}{\phi}_j  \label{work},
\end{equation} 
where $\Phi_j=\{\phi_1(\phi_j),\phi_2(\phi_j),..,\phi_j,...,\phi_N(\phi_j)\}$ is the magnetic configuration, for a given $\phi_j$, defined by the condition $T_i=0$ for every $i\ne j$.

\subsection{Transport Method}

The Keldysh non-equilibrium Green's function (NEGF) method is used here for describing the transport \cite{alex,alex2}. We expand the Hamiltonian $H(\{\phi\})$ and the one-electron wavefunctions in a localized atomic orbital basis set, and we decompose our system into two current/voltage probes sandwiching a central region. For an open current-carrying system $H$ can be written as
\begin{equation} \label{Ham2}
H=H_\mathrm{L}+H_\mathrm{R}+H_\mathrm{C}+H_\mathrm{LC}+H_\mathrm{RC}\;,
\end{equation}
where we have introduced the Hamiltonian for the left- (right-) hand side current/voltage probe $H_\mathrm{L}=H_\mathrm{L}(\{\phi\})$ ($H_\mathrm{R}=H_\mathrm{R}(\{\phi\})$), that for a central scattering region $H_\mathrm{C}=H_\mathrm{C}(\{\phi\})$, and the coupling matrix between the left (right) contact and the scattering region $H_\mathrm{LC}=H_\mathrm{LC}(\{\phi\})$ ($H_\mathrm{RC}=H_\mathrm{RC}(\{\phi\})$). The latter are indeed spin-polarized operators, i.e. $H=\sum_{\sigma} H^{\sigma}$, but for the sake of simplicity of the expression we skip the index $\sigma$ in the following formulas, thus refering to either of the spin components.

The NEGF method allows us to map this in principle infinite problem ($H_\mathrm{R}$ and $H_\mathrm{L}$ are infinite matrices) on an auxiliary finite problem. The key observation is that one can describe the effects of the current/voltage probes over the scattering region by means of their corresponding self-energies $\Sigma_\mathrm{L}$ and $\Sigma_\mathrm{R}$. These are non-hermitian matrices which contain all information about the electronic structure of the probes and their occupation. They can be written as
\begin{equation} \label{sigmaL}
\Sigma_\mathrm{L}=H_\mathrm{LC}^\dagger{g}_{\mathrm{L}}H_\mathrm{LC}\;, \quad  \mathrm{and} \quad \Sigma_\mathrm{R}=H_\mathrm{RC}{g}_{\mathrm{R}}H_\mathrm{RC}^\dagger\;,
\end{equation}
where we have defined the surface Green's function (GF) for the left- (right-) hand side probe ${g}_{\mathrm{R}}$ (${g}_{\mathrm{L}}$). Hence the ``effective Hamiltonian" of the scattering region in the presence of the current/voltage electrodes is written as
\begin{equation}\label{effH}
H_\mathrm{eff}=H_{\mathrm{C}}+{\Sigma}_\mathrm{L}+{\Sigma}_\mathrm{R}\:.
\end{equation}
Note that this is a finite non-hermitian matrix. Consequently the number of electrons in the scattering region is not conserved. 

Now we can construct the retarded GF associated with the scattering region plus the leads
\begin{equation}\label{green}
{G}(E)=\lim_{\zeta\rightarrow 0^{+}} \left[\left(E+i\zeta\right)-H_{\mathrm{C}}-{\Sigma}_\mathrm{L}-
{\Sigma}_\mathrm{R}\right]^{-1},
\end{equation}
and the associated (non-equilibrium) density matrix
\begin{equation}
\label{den_mat}
\rho=\frac{1}{\pi}\int_{-\infty}^{\infty} dE \left( n_L(E)\eta^L_F(E)+n_R(E)\eta^R_F(E) \right) \;,
\end{equation} 
where $n_{A}(E)=G(E)\Gamma_{A}(E)G^{\dagger}(E)$ are the partial density of state operators for electrons originating from each lead ($A=R,L$), $\Gamma=i (\Sigma(E)-\Sigma^{\dagger}(E))/2$ is the non-Hermitian part of the self-energy, $\eta^{A}_F(E)=\eta_F(E-\mu_{A})$ are the corresponding Fermi distribution function for the electron reservoirs with chemical potential $\mu_{A}$. 

Under our basic assumption of ``reflectionless" leads we can decouple the subsystems of electrons originating from the left and the right lead and treat them as separate statistical systems with electrochemical potentials $\mu_{L(R)}$. In equilibrium $\mu_L=\mu_R=\mu$, and a finite bias $V$ is introduced as $\mu_{L(R)}=\mu\pm |e|V/2$, so that the electron flow is from the left to the right lead for $V>0$. Thus the bias $V$ is assumed not to change the electronic structure of the leads, but only to rescale the energy levels. In practice $V$ is introduced as a rigid shift of lead-Hamiltonian on-site energies
\begin{equation}
{H}_\mathrm{L/R}\longrightarrow {H}_\mathrm{L/R} \pm \frac{V}{2} {\cal I},
\end{equation}
where ${\cal I}$ is the identity matrix for the respective lead. 

Self-consistency in our calculation is introduced by assuming that for a given magnetic configuration 
$\{\phi\}$ the Hamiltonian $H_\mathrm{C}$ depends solely on the scattering region density matrix 
$H_\mathrm{C}=H_\mathrm{C}[\rho]$. This is equivalent to assuming that the underlying electronic structure 
theory is a density-based theory, such as Hartree-Fock or density functional theory. In this case the set 
of equations (\ref{effH}), (\ref{green}) and (\ref{den_mat}) 
defines the self-consistent procedure. First one computes the scattering region GF (equation 
(\ref{green})) for $H_{\mathrm{C}}[\rho_0]$ 
evaluated at an initial density matrix $\rho_0$. Then from the GF a new charge density $\rho_1$ is calculated and used 
to construct the new Hamiltonian $H_{\mathrm{C}}[\rho_1]$. This procedure is iterated until reaching self-consistency, 
that is, until $\rho_{n+1}=\rho_n$. 

Finally from the converged GF the net current is calculated as \cite{alex,alex2}
\begin{eqnarray}
I(V)=\frac{e}{h} \sum_{\sigma} \int_{-\infty}^{\infty} dE \,
\left( \eta^L_F(E)-\eta^R_F(E) \right) & \nonumber \\ \mathrm{Tr} \left( \Gamma_L(E) G^{\dagger} \Gamma_R(E) G(E) \right)^{(\sigma)} &
\end{eqnarray} 
where we have summed over the spin index $\sigma$.

\subsection{The Model}

The techniques described in the previous sections are general and can be applied to a large class of Hamiltonians. In this work we focus our attention on a simplified model, which contains the fundamental ingredients for describing a current-carrying magnetic point contact, but at the same time does not present massive computational overheads. The structure we investigate is schematically represented in figure \ref{contact}. It consists of two semi-infinite leads with a simple cubic lattice structure and a $3\times3$-atom cross section connected through a linear chain of three atoms. Each atom carries a local magnetic moment, our classical quantities, arising from the deeply localized \textit{d}-electrons. The magnetic configuration of the leads is fully polarized (all MMs in a given lead point in the same direction) and we investigate the situation where the magnetizations of the two leads are opposite to each other. In contrast the three MMs of the atoms in the chain are allowed to rotate. A given magnetic configuration of the chain is thus described by the three angular coordinates $(\phi_1,\phi_2,\phi_3)$, with the convention that $\phi=0$ ($\phi=\pi$) for a spin alignment parallel to that of the magnetic moment of the left (right) lead. We consider spin rotations only in the $x$-$y$ plane thus neglecting the longitudinal angle (as in a Bloch wall). The alternative choice is to consider MM rotations in the $x-z$ plane  (a Neel wall), but as far as we neglect the magnetocrystalline anisotropy, these two models are identical. 

\myFig{1.0}{0.5}{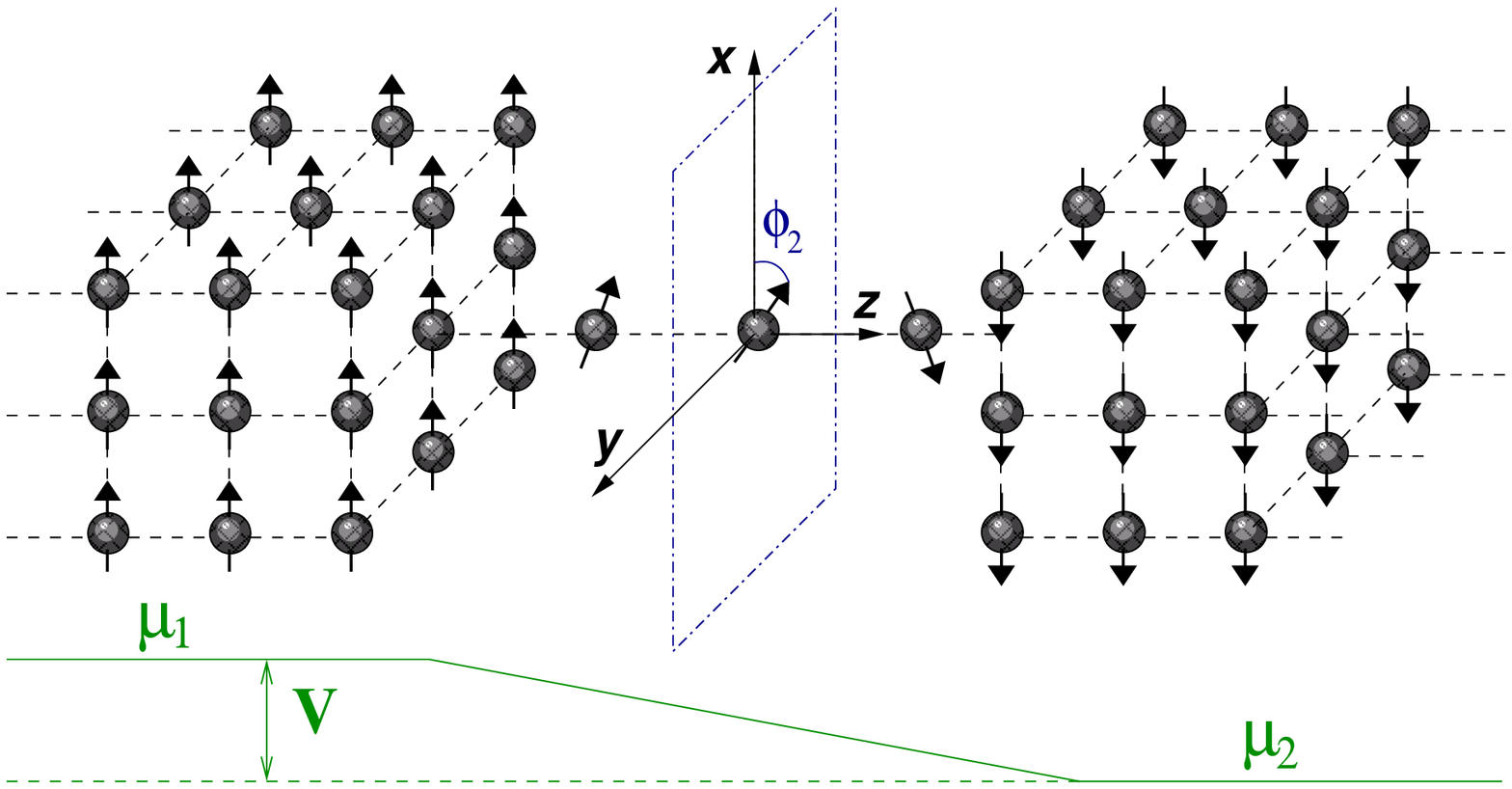}{Scheme of the magnetic point contact}{contact}

The current is carried by electrons belonging to an $s$ band, which is described by means of a single-orbital 
(plus spin) TB model. The Hamiltonian of equation (\ref{Ham}) is therefore explicitly written as
\begin{equation} \label{HamTB}
H(\{\phi\})=\sum_{i,j}[(H_\mathrm{e})_{ij}+(V_\mathrm{int})_{ij}]\,c^\dagger_i c_j + V_{{\rm classical}}(\{\phi\})\;,
\end{equation}
where $c^\dagger_i$ and $c_j$ are creation and annihilation operators for electrons at the atomic 
sites $i$ and $j$ respectively. The matrix elements of the ``free'' electron part are those of a 
nearest-neighbor TB model
\begin{equation}
(H_\mathrm{e})_{ij}=[\epsilon_0+U(\rho_i-\rho_i^0)]\delta_{ij}+\gamma\delta_{i,j\pm 1}\;,
\end{equation}  
where $\epsilon_0$ is the on-site energy, $\gamma$ is the hopping parameter, $U$ is the on-site Coulomb 
repulsion, $\rho^0$ the reference on-site charge corresponding to the neutral free atom, and $\rho$ is 
the self-consistent local charge. The potential $V_\mathrm{es}$ of equation (\ref{Ham}) has now
been separated in two parts: $V_\mathrm{int}$ and $V_\mathrm{classical}$. The interaction between conduction 
electrons and the local MMs is contained in $V_\mathrm{int}$ which in our model reads
\begin{equation}\label{Ves}
(V_\mathrm{int})_{ij}=-s_i\frac{J}{2}\cos{\phi_i}\,\delta_{ij}\;.
\end{equation}  
where $s_i=\rho_i^\uparrow-\rho_i^\downarrow$ is the local spin polarization of the $s$ electrons at 
site $i$ and $J$ is the exchange parameter. Therefore $V_\mathrm{int}$ describes a Heisenberg-type interaction 
between the local classical MMs and the current carrying $s$ electrons. Finally, the classical term $V_{{\rm classical}}$ 
parameterizes the interaction between local MMs. Here we assume a Heisenberg spin-spin interaction
\begin{equation}
V_\mathrm{classical}=-\frac{J_{dd}}{2}\sum_{i,j}\pmb{S}_i\cdot\pmb{S}_j=-\frac{J_{dd}}{2}\sum_{i,j}\cos{(\phi_i-\phi_j)}\:,
\end{equation}  
where $J_{dd}$ is the intersite exchange integral and we have assumed normalized classical spin $|\pmb{S}_i|=1$, in such a way that $|\pmb{S}_i|$ is incorporated in the definitions of $J$ and $J_{dd}$. In summary our model is that of $s$ conduction electrons exchanged coupled to local magnetic moments, in turn described by a Heisenberg-type energy. This is usually known as the $s$-$d$ model \cite{mota}.

The torque experienced by the $i$-th local MM in the chain is then obtained from equation (\ref{hft}) and reads
\begin{equation}
\label{trqs}
T_i= -\frac{J}{2} s_i \sin\phi_i-\frac{J_{dd}}{2} (\sin(\phi_i-\phi_{i-1}) +\sin(\phi_{i+1}-\phi_i))
\end{equation}  
where $i=1,2,3$, and we have defined $\phi_0\equiv0$, $\phi_4\equiv\pi$ since the magnetization of the two leads is considered pinned in an antiparallel arrangement.   

In this simple model the surface GF (at a general complex energy $E$) of the leads have an analytical form. In reciprocal space,
\begin{eqnarray}
\label{sGF}
& g(E,\pmb{k})=\frac{E-\epsilon(\pmb{k})-\sqrt{(E-\epsilon(\pmb{k}))^2-4\gamma^2}} {2\gamma^2},\quad \rm{Im} (E) >0 
\end{eqnarray} 
where $\epsilon(\pmb{k})$ is the energy, as a function of transverse wavevector (in appropriate unites) ${\bf k}=(k_{x},k_{y})$ with $k_{x} = 1,...,N_{x}$, $k_{y} = 1,...,N_{y}$ for an $N_x\times N_y$-atom simple-square monatomic slab in a nearest-neighbor orthogonal TB $s$-band model, 
\begin{equation} \label{sGF_E}
\epsilon(\pmb{k})\!=\!\epsilon_0-2|\gamma|\cos{\left(\frac{k_x\pi}{N_x+1}\right)} - 2|\gamma| \cos{\left(\frac{k_y\pi}{N_y+1} \right)}\;.
\end{equation} 
The expression of equation (\ref{sGF}) is then expanded over the real-space basis \cite{ch_GF} and used in the matrix equation for the self-energies. The definition of the complex square-root is given in \cite{pmb}.

\section{Results}

Here we investigate the magneto-dynamics of atomic point contacts, and in particular of the model structure described in figure \ref{contact}. The TB parameters are $\epsilon_0=-3$\,eV, $\gamma=-1$\,eV, $U=12$\,eV, which gives a large bandwidth for the $s$ electrons and provides local charge neutrality as expected in a metal. For the exchange parameters we investigate the range $0\leq J \leq 3$~eV and $0\leq J_{dd}\leq 5$~eV. However we have identified the values $J=1$\,eV and $J_{dd}=50$\,meV as a realistic choice for simulating the main physics of magnetic transition metals \cite{rushbrooke,hirsch,mota}, and we will refer to those values as the ``realistic parameters''.

We start our analysis with studying DW migration in the three atom chain. As in  figure \ref{contact}, the magnetic moments of the leads are in the antiparallel configuration, so that that a DW nucleates in the chain. We then investigate the displacement of the Bloch wall from the interface between the first and the second atom in the chain to that between the second and the third (generated by a rotation of the magnetic moment of the middle atom). These simulations use the realistic parameters given above, so they can be related to point contact experiments \cite{viret}.  Both the cases of spatially symmetric and asymmetric chains are studied. Then we explore the effect of varying the strength of the exchange parameters, and we identify three different regimes. Finally, we revisit the problem of whether or not generalized forces away from equilibrium are conservative, and demonstrate numerically that the torques in the present system under current flow are not conservative.

\subsection{Domain wall migration}

By performing numerical minimization of all the torques, exerted on the MM in the constriction, with various initial conditions, we have determined that all eight collinear arrangements, such as $(0,0,0)$, $(0,0,\pi)$, $(0,\pi,\pi)$, $(\pi,\pi,0)$ etc. are stable zero-torque magnetic configurations and we have studied various transitions between them (see fig.\ref{bigJ}). In particular, we have investigated in detail the migration of an abrupt DW within the atomic chain, i.e. the transition between the $(0,0,\pi)$ and $(0,\pi,\pi)$ magnetic configurations, achieved by rotation of $\pmb{M}_2$ as described above. Physical characteristics of this process as function of $\phi_2$ are presented in Fig.\ref{T2_ex}. It is observed that during the rotation of $\pmb{M}_2$ its neighboring MMs experience small tilts from the collinear alignment and after a turning point fall back to their initial state. The intersite exchange coupling is not strong enough to induce spin flips of the neighboring MM along with the one that is rotated and even hypothetical values of $J_{dd}$ up to 0.4\,eV do not change this picture (see fig.\ref{effJd}). This observation suggests that the dynamical processes of the MMs in the constriction can be decomposed into series of single MM rotations.     

\myFig{1.0}{0.7}{figure02.eps}{\footnotesize{Typical calculation of microscopic properties during DW migration in the contact (for $J=1$eV, $J_{dd}=50$meV) as function of the ``reaction coordinate" $\phi_2$: (a) The stable angular variables $\phi_1$ and $\phi_3$; (b) The three on-site spin polarizations $s_i=\rho_i^{\uparrow}-\rho_i^{\downarrow}$; (c) Torque and work, performed by the MM; (d) Net current at $V=0.5$\,V. The voltage in panels (a)-(c) is zero.}}{T2_ex}

The torque $T_2$, computed as a function of $\phi_2$ (fig.\ref{T2_ex}c) at every point on the way, is interpolated and integrated according to (\ref{dW}) to determine the effective energy barrier for the DW migration 
\begin{equation} 
W(\phi_2)=-\int_0^{\phi_2} T_2 \,\rm{d}\phi_{2}'\:.  \label{work2}
\end{equation} 

Because of the specific geometric and time-reversal symmetries of the system, the two states $(0,0,\pi)$ and $(0,\pi,\pi)$ are macroscopically identical. Thus the calculated energy barrier between them is symmetric and the total work $W(\pi)$ for the quasi-static process is zero (fig.\ref{T2_ex}c). The activation energy for this process in our TB parameterization is $54$\,meV. It is found that the conductance of our system depends on the alignment of the MM and in this case the net current shows a symmetric bell-shaped dependence on $\phi_2$ (fig.\ref{T2_ex}d). For this case ($V=0.5$\,V), the conductance varies from 1.76$e^2/h$ at the collinear states $\phi_2=0,\pi$ to a maximum of 1.86$e^2/h$, reached at $\phi_2=\pi/2$. 

Further, it is observed that the external bias, driving a spin-polarized current, suppresses the response of $\phi_{1,3}$ to the motion of $\phi_{2}$ (fig.\ref{effV}a,b) but enhances the onsite polarizations (fig.\ref{effV}c) as well as the energy barrier (fig.\ref{effV}d). At any finite temperature, this phenomenon would manifest itself as suppression, with increasing bias, of the frequency of DW transitions back and forth between the two stable magnetic configurations. The net current profile is slightly sharpened as the bias increases (fig.\ref{effV}e) and it also becomes more spin-polarized due to the increased misalignment of the correspondent spin-polarized bands in the two leads (fig.\ref{effV}f).

\myFig{1.0}{0.8}{figure03.eps}{\footnotesize{Effect of the external bias on the microscopic properties of the contact as function of $\phi_2$: the zero-torque positions of (a) $\phi_1$ and (b) $\phi_3$;  (c) the on-site spin polarizations $\{s_i= \rho_i^{\uparrow}- \rho_i^{\downarrow}\}_{i=1,2,3}$;  (d) the work profile; (e) the total net current and (f) its polarization $\kappa_I=(I_{\uparrow}-I_{\downarrow})/ (I_{\uparrow}+I_{\downarrow})$}}{effV}

\subsection{Non-uniform contact}

Current-induced relaxation of the atomic positions can break the spatial symmetry in point contacts similar to 
ours \cite{ch_frc} and substantially weaken the stability of these systems. To investigate the effect of a small 
asymmetry in the contact geometry on the migration barrier for the DW, we map the displacement of the middle atom 
from its symmetric position onto a small variation of hopping integrals between the middle atom and its neighbors 
in the chain
\begin{equation}
\gamma_{12}=\gamma(1+\delta), \quad \gamma_{23}=\gamma(1-\delta)\:.
\end{equation}  

It results in breaking the symmetry about $\phi_2=\pi/2$ of the effective energy barrier observed in all our previous calculations (fig.\ref{asym_bar}). The total work for the $(0,0,\pi) \rightarrow (0,\pi,\pi)$ transition is negative thus the internal energy of the classical MM is increased. The degeneracy of the $(0,0,\pi)$ and $(0,\pi,\pi)$ state is lifted, as the spatial symmetry, associated with a reflection plane at $z=0$, is no longer present. When the hopping parameters are altered by $\delta=5\%$, as if the middle atom is slightly shifted to the left, the $(0,\pi,\pi)$ configuration becomes energetically preferable, alternatively, $\delta=-5\%$ favors the $(0,0,\pi)$ state, with all the microscopic properties being invariant to a change $\phi_2\rightarrow(\pi-\phi_2)$ (fig.\ref{asym_bar}).

\myFig{1.0}{0.4}{figure04.eps}{\footnotesize{Effect of an asymmetry in the hopping integrals in the 3-atom chain with $\delta=\pm 0.05$ on the barriers for $(0,0,\pi)\rightarrow(0,\pi,\pi)$ transition at different bias: (a) $V=0$\,V; (a) $V=1$\,V(a); $V=2$\,V.}}{asym_bar}

We have presented the typical microscopic properties in figure \ref{asym_all}. As expected the effective ferromagnetic coupling between the MMs is further increased by the enhanced electronic coupling. The onsite polarizations of all the atoms shift almost rigidly as the middle atom is brought towards one or the other of the leads. The net current shows significant asymmetry from the regular bell-shaped dependence on $\phi_2$ and the more stable configuration is always found to be less conducting (fig.\ref{asym_all}d). The I-V characteristics of the previously degenerate $(0,0,\pi)$ and $(0,\pi,\pi)$ states is split into two branches, whose displacement increases with voltage (fig.\ref{asym_all}f) and reaches 10\% for $V=2$\,V. Thus we expect DW migrations within the constriction, in the case of small deviations from a uniform geometry, to be accompanied by random-telegraph-noise-like variations in the net current. The interplay between the current-induced relaxation of the magnetic and mechanical degrees of freedom is the subject of work in progress.   

\myFig{1.0}{0.7}{figure05.eps}{\footnotesize{Effect of an asymmetry in the hopping integrals with $\delta=\pm 5\%$ on the microscopic properties of the contact (as in fig.\ref{effV})}}{asym_all}

\subsection{Mapping out the parameter space}

The torques defined in (\ref{trqs}) depend explicitly on the exchange parameters $J, \,J_{dd}$ and the balance of these two coupling mechanisms determines the spin dynamics in the constriction. Figure \ref{effJd}(a,b,c) describes the variation of the equilibrium magnetic properties of the $(0,0,\pi) \rightarrow (0,\pi,\pi)$ transition as the intersite exchange strength $J_{dd}$ is varied. The effect of greater $J_{dd}$ is increased disalignments of the two neighboring MM $\phi_{1,3}$ following the rotation of $\phi_2$ (\ref{effJd}a,b). Because of this the corresponding spin-polarizations $s_{1,3}$ (but not $s_2$) are slightly affected (\ref{effJd}c) as a result of the stronger coupling. The typical bell-shape of the current vs. $\phi_2$ is broadened as $J_{dd}$ increases (fig.\ref{effJd}e) due to the fact that stronger intersite exchange coupling tends to make the three MMs in the contact more uniformly distributed in angle, which makes the contact better-conducting for any $\phi_2$. Also the current is somewhat less polarized in the case of stronger intersite coupling (fig.\ref{effJd}f).

\myFig{1.0}{0.7}{figure06.eps}{\footnotesize{Effect of the strength of the intersite coupling $J_{dd}$ on the microscopic properties (as in fig.\ref{effV}) in equilibrium (a,b,c,d) and at $V=1$\,V}}{effJd}

The graph in fig.\ref{Jdd_trq} describes the distribution of stability patterns of the three MMs in the constriction as a function of the exchange parameters. It is based on a study of the DW migration barrier for different exchange parameters. We can distinguish three different regimes depending on the values of the exchange parameters: (1) ``magnetostatic regime" associated with the presense of 2 stable magnetic states, for which $\phi_2=\pm\pi/2$; (2) ``mixed regime": 4 stable configurations, 2 in each half-plane for which $0<\phi_2<\pi/2$ and $\pi/2<\phi_2<\pi$; (3) ``current-driven regime": 8 stable configurations, namely all the collinear MM alignments $(0,0,0)$, $(\pi,\pi,\pi)$, $(0,0,\pi)$, $(0,\pi,\pi)$, $(\pi,0,0)$, $(\pi,\pi,0)$, $(0,\pi,0)$, $(\pi,0,\pi)$ (fig.\ref{bigJ}). The last case is confirmed by full torque minimizations at various initial conditions. It should be noted that stability of the two unipolarized collinear configurations is also found above the boundary on figure \ref{Jdd_trq} but they are then only accessible once the system is trapped within very narrow regions of $\{\phi\}$ space. Our main observation is that the collinear configurations are the only form of stability of the magnetic chain in the contact for a range of exchange parameters around the ``realistic'' values, defined earlier. Thus, even though our calculations in the previous and in the next section use $J_{dd}=50$\,meV and $J=1$\,eV, the qualitative features of the results may be expected to hold for a range of values of the parameters, with $J_{dd}$ in the region of tens of meV and $J\gtrsim0.5$\,eV.

\myFig{0.9}{0.8}{figure07.eps}{\footnotesize{Diagram showing the three regions, with 2, 4 and 8 stable magnetic configurations respectively (see text), in a $J$-$J_{dd}$ cut of the parameter space. The boundaries between the three regimes are calculated from the properties of the $(0,0,\pi) \rightarrow (0,\pi,\pi)$ transition (see inset). The dashed lines correspond to voltages of 1 and 2 V.}}{Jdd_trq}

It is observed that the bias, driving a spin-polarized current, is able to distort the boundaries between these three regions significantly (fig.\ref{Jdd_trq}, dashed lines refer to $V=1,2$ V). The effect of the bias on the migration energy-barrier profile has been studied for different values of the exchange parameters and reported for two representative cases $J=1$ eV and $J=2.5$ eV (fig.\ref{biasJ1}). The overall observation is that for $J\lesssim1.5$ eV (the ``realistic'' regime) increasing bias (current) enforces the barrier, while for $J\gtrsim2.5$ eV the bias suppresses the barrier. For intermediate values of $J$ the barrier shows a non-monotonic behavior with bias.

\myFig{1.0}{0.7}{figure08.eps}{\footnotesize{The biased energy barriers for the transition $(0,0,\pi)\rightarrow(0,\pi,\pi)$ at different voltages V=0,1,2\,V, $J=1.0$\,eV (left) and $J=2.5$\,eV (right) and for three values of $J_{dd}$ (one in each of the ranges, discussed in the text): (a) $J_{dd}=12$\,meV; (b) $J_{dd}=0.25$\,eV; (c) $J_{dd}=0.3$\,eV}; (d) $J_{dd}=1$\,eV; (e) $J_{dd}=2$\,eV; (f) $J_{dd}=3$\,eV.}{biasJ1}

\subsection{Current-voltage characteristics}
 
The current-voltage characteristics of the system at all the different stable alignments of the MM in the chain are presented in figure \ref{IVs}. All I-V curves are symmetric about the origin, due to the time reversal symmetry. The eight possible stable magnetic states are $4\times2$ degenerate with respect to the current, which, as discussed above, is due to the specific spatial symmetry of the atomic contact. The slopes of the I-V curves cannot be directly related to the number of DWs in the constriction (three for the $\{0,\pi,0\}$ and $\{\pi,0,0\}$; one for $\{0,0,0\}$ and $\{0,0,\pi\}$ states). Nevertheless, the least steep curve does correspond to the highest number of domain walls present within the chain. As soon as the DW migrates from the chain towards the leads (as in the $\{0,\pi,0\}$ to $\{\pi,0,0\}$ transition) the conductance increases. We conclude that microscopic magnetization reversals in the constriction could be causing massive current variations (of up to 50\%) at a given bias. At a given finite temperature this would result in a random telegraph noise in conductance measurements and such effects have been observed experimentally \cite{pufall,urazhdin,viret}. 

Figure \ref{bigJ} represents the work for successive 1 MM rotations. The sequence of transitions goes through every stable magnetic configuration once, and returns to the initial state. In equilibrium the depths of the wells in this graph correspond to the relative energies of our system in various stable magnetic states with respect to the initial one. Thus $(0,\pi,0)$ and $(\pi,0,\pi)$, which have two abrupt DW within the chain and are the least conducting states, are found to be the most stable among the collinear alignments. It is observed that the external bias has a non-trivial effect on the effective energy barriers for these transitions. The total work for the closed loop cancels out for any bias. This, however, is not an indication of  conservativeness of the current-induced torques (\ref{hft}), but is rather an artefact of the specific symmetries in this certain closed path, which includes all the states and can be decomposed into two branches going through identical states in opposite direction. 

There are $12=(5\times2+2)$ one-MM-rotation transitions between different pairs of stable collinear alignments, out of which we distinguish $5+2=7$ different transitions. The average activation barrier at equilibrium is 71\,meV with a variance of 36\,meV and it depends non-monotonically on the bias: 65.6\,meV at 1\,V and 68\,meV at 2\,V. These values for the activation barriers suggest switching frequencies, and hence random telegraph noise in the current, in the microwave range at room temperature.

\myFig{0.9}{0.7}{figure09.eps}{\footnotesize{The current-voltage characteristics at different stable collinear alignments for the MM in the chain. The inset represents the correspondent spin-polarization of the net current $\kappa_I=(I_{\uparrow}-I_{\downarrow})/ (I_{\uparrow}+I_{\downarrow})$, calculated for each state. It is a non-monotonic function of the bias.}}{IVs}

\myFig{1.0}{0.4}{figure10.eps}{\footnotesize{Work for successive transitions between the 8 stable magnetic configurations at different bias $V=0,1,2$\,V.}}{bigJ}

\subsection{Are the torques conservative?}

The question if, and under what conditions, forces under steady-state current are conservative remains an 
open fundamental problem in the theory of transport \cite{tm}. A thermodynamic formulation of forces under 
non-equilibrium steady-state conditions, proposed in reference \cite{pmb}, leads to the explicit identification 
of a thermodynamic potential for electro-migration \cite{mr}. However, as a consequence of the infinite nature 
of open-boundary systems, this potential involves a conditionally convergent real-space summation. If the sequence 
of terms in this summation remains invariant along a given path in the configuration space of the system, then along 
that path, current-induced generalized forces are rigorously expressible as gradients of a scalar potential and are 
therefore conservative. The possibility remains open, however, that the order of terms in the conditionally 
convergent sum may change, as specific points, or manifolds, in configuration space are traversed \cite{mr}. 
This constitutes an effective breakdown of the Born-Oppenheimer approximation, with the consequence that paths 
that span such points are non-conservative \cite{mr}. 

We now carry out a numerical test to see whether or not the torques in equation (\ref{trqs}) are conservative 
in the present current-carrying system. The work for a set of transitions between collinear MM configurations, 
performed by rotation of a single MM, which form a closed-loop, is calculated for different voltages. The 
full work for three different loops of four consecutive transitions as a function of the applied voltage is 
presented in figure~\ref{loopV}. A significant variation of the closed-loop work with bias is observed.

In order to resolve the numerical error we have performed a series of tests with different levels of accuracy. We recognize  several sources of numerical error: (1) The level of convergence of the density matrix $\delta\rho$. (2) The fineness of the energy mesh for the contour integration in the complex plane $\delta E$. (3) The level of torque relaxation $\delta T$. (4) The angular mesh for the torque integration which results into the work. As the torque in the current-driven regime is a very smooth function of the reaction coordinate (fig.\ref{T2_ex}c), we have found (4) insignificant for the value of the integral (\ref{work}). The effect of the rest of the accuracy parameters on the closed loop work (see the loop $\{\pi\pi\pi-\pi\pi0-0\pi0-0\pi\pi-\pi\pi\pi\}$ in fig.\ref{loopV}) is summarized in table.\ref{acctab}. 

\begin{table}[htp]
\begin{ruledtabular}
\begin{tabular}{cddddd}
\text{\small($\delta\rho$,$\delta E$,$\delta T$)}, \%  &  \multicolumn{1}{c}{$W_{0\,\rm{V}}$}  & \multicolumn{1}{c}{$W_{0.5\,\rm{V}}$}  & \multicolumn{1}{c}{$W_{1\,\rm{V}}$}  & \multicolumn{1}{c}{$W_{1.5\,\rm{V}}$} & \multicolumn{1}{c}{$W_{2\,\rm{V}}$} \\ \colrule \\
\text{\small(100,100,100)} & 0.0001 & -1.407 & -7.454 & 0.601 & 19.81 \\
\text{\small(100,20,100)} &  0.0001 & -1.290 & -7.361 & -0.036 & 19.79 \\
\text{\small(1,100,2)} &  0.0406 & -1.300 & -7.356 & 0.662 & 19.88 \\
\end{tabular}
\end{ruledtabular}
\caption{\footnotesize{The work (in meV) for the loop $\{\pi\pi\pi-\pi\pi0-0\pi0-0\pi\pi-\pi\pi\pi\}$ as function of the accuracy parameters (in relative units). The value of the bias V is given as a subscript.}} \label{acctab} 
\end{table}     

The results in table~\ref{acctab} suggest that the effect of nonzero work for a closed-loop set of transitions is not substantially affected by variations of 1--2 orders of magnitude about the chosen level of accuracy. Thus we see that in the present case, along the selected closed paths, we have an explicit example of non-conservative generalized non-equilibrium forces. 

\myFig{0.9}{0.7}{figure11.eps}{\footnotesize{Dependence of closed-loop work on the voltage}}{loopV}

\section{Conclusions}

In conclusion, we have proposed a microscopic quantum-classical approach for computing the current-induced torques on the local magnetization in ferromagnetic point contacts under bias. Our method employs an $s$-$d$ model for the electronic structure and NEGF technique for describing the electronic transport. The direction of the local MMs are mapped onto classical degrees of 
freedom. We apply this method to a specific atomic structure, which consists of a monoatomic chain, bridging over two semi-infinite leads with opposite magnetizations, so that at least one magnetic DW is formed within the constriction. We investigate the stability of various magnetic configurations, involving multiple DWs, and the effect of bias driving a spin-polarized current, on the 
energy-barrier for the DW migration. For realistic values of exchange parameters only the collinear MM arrangements are stable. These configurations carry different (by up to 50 \%) net currents, and the average activation barrier for transitions is about
65--70\,meV with variance of 20--40\,meV, depending on bias. Therefore random telegraph noise in current with significant amplitude could be related to thermally-activated MM rearrangements within the constriction. We have also found that geometrical asymmetries in the atomic structures (which could be induced by the current \cite{ch_frc}) affect the symmetry of the activation barrier for DW migration, pinning it to a preferential spatial position, in which case the structure becomes less conducting. 

The observation that the collinear MM alignments are the only stable magnetic states and the fact that direct intersite interaction is not able to induce flips in the neighboring MMs as one MM in the chain is quasi-statically rotated, has enabled us to calculate the work for series of successive magnetic rearrangements of MMs, involving single-MM rotations. Thus, we could address numerically the long-standing question of the conservativeness of the current-induced forces (torques, in our case) in open-boundary non-equilibrium system. We have found numerical evidence that the work for various closed-loop paths is not zero, but varies non-monotonically as the system is driven away from equilibrium. Hence generalized current-induced forces in our present system are not conservative, at least in the section of the configuration space spanned by the present calculations.

\begin{acknowledgments}
This work has been sponsored by the Irish Higher Educational Authority under the North South Programme for Collaborative Research. The author M. S. is grateful to A. R. Rocha for the transport code development and helpful discussions.  
\end{acknowledgments}

\end{document}